\documentclass[%
 aip,
 jmp,%
 amsmath,amssymb,
 reprint,%
]{revtex4-1}

\usepackage{graphicx}
\usepackage{dcolumn}
\usepackage{bm}
\usepackage{ae} 
\usepackage{epstopdf}

\begin{document}

\hyphenation{o-pe-ra-tions}

\hyphenation{me-tho-do-lo-gy}

\hyphenation{me-tho-ds}

\hyphenation{i-ma-ge}

 \preprint{AIP/123-QED}

\title[]{Total 3D imaging of phase objects using defocusing microscopy: application to red blood cells}

\author{P. M. S. Roma$^{1}$, L. Siman$^{1}$, F. T. Amaral$^{2}$, U. Agero$^{1}$ and O. N. Mesquita}
\email[E-mail me at: ]{mesquita@fisica.ufmg.br}
\address{$^{1}$Departamento de F\'\i sica, $^{2}$Programa de P\'os Gradua\c{c}\~ao em
Engenharia El\'etrica, Universidade Federal de Minas Gerais, Caixa
Postal 702, CEP 31270-901, Belo Horizonte, MG, Brazil.}

\begin{abstract}
We present Defocusing Microscopy (DM), a bright-field optical
microscopy technique able to perform total $3$D imaging of
transparent objects. By total 3D imaging we mean the determination
of the actual shapes of the upper and lower surfaces of a phase
object. We propose a new methodology using DM and apply it to red
blood cells subject to different osmolality conditions: hypotonic,
isotonic and hypertonic solutions. For each situation the shape of
the upper and lower cell surface-membranes (lipid
bilayer/cytoskeleton) are completely recovered, displaying the
deformation of RBCs surfaces due to adhesion on the glass-substrate.
The axial resolution of our technique allowed us to image
surface-membranes separated by distances as small as $300$ nm.
Finally, we determine volume, superficial area, sphericity index and
RBCs refractive index for each osmotic condition.

\end{abstract}

\maketitle

Standard optical microscopy imaging of phase objects is generally
obtained with microscopes operating in Phase Contrast or Nomarsky
configurations\cite{zernicke,nomarski}. However, even for objects
with uniform refractive index these techniques present difficulties
for obtaining accurate full-field thickness profiles. Recently, new
approaches known as quantitative phase microscopy techniques
\cite{nugent1,nugent2,rappaz2008,popescu2008,kononenko}, have
adequately obtained full-field height profile of phase objects with
successful application in living cells. Despite that, in the case of
red blood cells (RBCs) these techniques are only able to provide the
thickness profile and thickness amplitude fluctuations, such that
height profile information of each particular cell surface-membrane
(lipid bilayer/cytoskeleton) is not accessed.

Using Defocusing Microscopy (DM), we present a new method able to
perform total $3$D imaging of RBC, and thus, capable of determining
the actual shape of the cell's upper and lower surface-membranes,
separately. New procedures to retrieve RBC volume, superficial area,
sphericity index and refractive index using DM are also exposed. All
developed methods are applied to data of $25$ RBCs immersed in three
different solutions: hypotonic (200\;mOsm/kg), isotonic
(300\;mOsm/kg) and hypertonic (400\;mOsm/kg), showing the
differences in the lower membrane adhesion to the glass substrate.
For assessment of bio-mechanical properties along the RBCs surfaces,
nanometer height fluctuations for each interface can be obtained
separately, such that the effect of adhesion to the substrate can
also be evaluated \cite{giuseppe,livia}. A defocusing technique has
been recently applied for $3$D imaging of cells using a phase
contrast microscope under white-light illumination, with transverse
resolution of $350$ nm and axial resolution of $900$ nm
\cite{popescunature}. This technique cannot resolve
surface-membranes separated by an axial distance smaller than $900$
nm, which is the case of most RBCs. Strikingly, our Defocusing
Microscopy technique, using a bright-field defocused microscope and
our theoretical framework, can resolve surface-membranes of RBCs
separated by axial distances down to $300$ nm.

Transparent objects that would be invisible in a standard
bright-field optical microscope can turn visible by defocusing the
microscope, since the act of defocusing introduces a phase
difference between the diffracted orders and the non-diffracted
transmitted order (zero order). From contrast intensity measurements
of images in two focal positions, one can obtain information about
the phase of the optical electric field and reconstruct the height
profile of the phase object. The formalism of Intensity Equation has
been used for this purpose \cite{teague,nugent1}, but in its present
form there are no explicit phase terms taking into account the
distance between the objective focal plane and the diffracting
surfaces, such that the total $3$D imaging is not feasible. In our
approach, Fresnel Diffraction Theory and the formalism of
propagation of the Angular Spectrum\cite{wolf,goodman} are used in
order to propagate the light electric field throughout our defocused
microscope. We treat the electric field as a scalar quantity and
neglect polarization effects. As a result, the defocused electric
field is represented by an integral over the diffracted wave-vectors
($\vec{q}$), where the positions of the objective focal plane
$(z_{f})$ and the phase object surfaces appear explicitly, allowing
us to obtain total $3$D imaging of phase objects. Our model works
for pure phase objects, such that in the case of RBCs the
illumination light should be red filtered to minimize light
absorption. For the total $3$D imaging using DM, coherence of the
illumination light is not an important issue. More details of the
experimental set up are given in the supplementary material
\cite{sm}.

\begin{figure}
\centering\includegraphics[width=7cm]{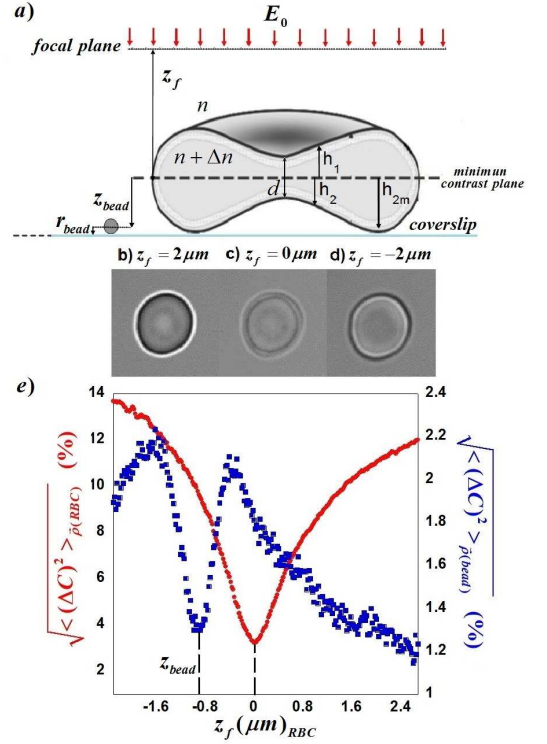}
\caption{\label{hemacia} (a) RBC model with the reference frame
origin at the cell minimum contrast plane, defining $h_{1}>0$ and
$h_{2}<0$ as the coordinates of the upper and lower
surface-membranes, respectively. (b-d) Contrast images of a RBC at
different defocusing distances $z_f$. (e) Contrast standard
deviation profile as a function of defocusing distance $z_f$ for a
RBC and the bead. The minimum values for both profiles are clearly
determined, and the distance between them is $|z_{bead}|$, such that
$|h_{2m}|=|z_{bead}|+|r_{bead}|=|z_{bead}|+0.099\mu m$.}
\end{figure}

In the model presented, all parameters associated to optical
elements that are displaced to produce image defocusing are known
\cite{livia}. In our derivation below, defocusing is caused by the
displacement of the microscope objective, which can be related to
the displacement of other elements, as will be shown later. The
results obtained are independent of the origin of the reference
frame along z, adopted here as the plane where the RBC contrast has
a minimum value, that corresponds to cell largest diameter plane.
The defocused light electric field that passes through a red blood
cell (Fig.\ref{hemacia}a) can be written as
\cite{leo,bira2003,coelho2005,coelho2007,giuseppe,livia},
\begin{eqnarray}\label{ap3g}
E(\vec{\rho},z_{f})=\frac{1}{(2\pi)^2}\displaystyle{\int_{-\infty}^{+\infty}}
[A_1(\vec{q})\:e^{i\vec{q}\cdot\vec{\rho}}\:e^{i(\frac{(z_f-h_1(\vec{\rho}))q^2}{2k})}+\nonumber\\
A_2(\vec{q})\:e^{i\vec{q}\cdot\vec{\rho}}\:e^{i(\frac{(z_f-h_2(\vec{\rho}))q^2}{2k})}]d\vec{q},\nonumber\\
\end{eqnarray}
where $h_1(\vec{\rho})$, $h_2(\vec{\rho})$ are the coordinates for
the upper and lower RBC surface-membranes, respectively,
$k=n_{ob}k_0$ is the light wavenumber, with $n_{ob}$ the objective
immersion medium index of refraction and $k_0$ the wavenumber in
vaccum. The terms
\begin{eqnarray}\label{ap4g}
A_{1}(\vec{q})=\displaystyle{\int_{-\infty}^{+\infty}}E_{0}\:e^{-i\vec{q}\cdot\vec{\rho}}\:e^{i{\Delta
nk_0h_1(\vec{\rho})}}\:d\vec{\rho}\nonumber\\
A_{2}(\vec{q})=\displaystyle{\int_{-\infty}^{+\infty}}E_{0}\:e^{-i\vec{q}\cdot\vec{\rho}}\:e^{-i{\Delta
n k_0h_2(\vec{\rho})}}\:d\vec{\rho}
\end{eqnarray}
are the Angular Spectra of the electric field diffracted by the
surface-membranes. Additionally, $\Delta n$ is the refractive index
difference between the RBC and its surrounding medium (uniform
inside the RBC), and $h_{1/2}(\vec{\rho},t)=h_{1/2}(\vec{\rho}) +
u(\vec{\rho},t)$, such that the time average\;\;
$<h_{1/2}(\vec{\rho},t)>=h_{1/2}(\vec{\rho})$ and $u(\vec{\rho},t)$
is the out-of-plane height fluctuations in the nanometer range.
Since we are interested in equilibrium shapes of RBCs, Eqs.
(\ref{ap3g}) and (\ref{ap4g}) are time averaged quantities. The
second order cross terms $A_{1}(\vec{q}).A_{2}(\vec{q})$ are
neglected (in the worst case $(\Delta n k_0)^2 h_1 h_2$ $<1$), which
means that the incident light is diffracted by only one interface
and the chance of being diffracted by the two interfaces is small.
Naturally, higher order terms can be included into the model if they
become necessary for a better quality  $3$D image.

Defining the image contrast as
$C(\vec{\rho})=\frac{I(\vec{\rho})-I_0}{I_0}$, where $I(\vec{\rho})$
is the intensity of the object image  and $I_0$ is the background
intensity, for first-order diffraction, the DM contrast for a red
cell decomposed in spatial Fourier components $\vec{q}$ is,
\begin{eqnarray}\label{ap2g}
C(\vec{\rho})=\frac{2\Delta nk_{0}}{\sqrt{A}}\Bigg\{\sum_{\vec{q}}
\Bigg[h_2(\vec{q})\sin\Bigg(\frac{(z_f-h_2(\vec{\rho}))q^2}{2k}\Bigg) \nonumber\\
-h_1(\vec{q})\sin\Bigg(\frac{(z_f-h_1(\vec{\rho}))q^2}{2k}\Bigg)\Bigg]\sin(\vec{q}\cdot\vec{\rho})\Bigg\},
\end{eqnarray}
where $A$ is the RBC surface area and
$h(\vec{\rho})=\frac{1}{\sqrt{A}}{\sum_{\vec{q}}}
h(\vec{q})sin(\vec{q}\cdot\vec{\rho})$. For the chosen reference
frame, $h_1(\vec{\rho})
> 0$ and $h_2(\vec{\rho}) < 0$, but for convenience we will use
positive quantities, such that $h_2=-|h_2|$. For small defocusing
distances the approximation $sin
((z_{f}-h)q^2/2n_{ob}k_0)\simeq(z_{f}-h)q^2/2n_{ob}k_0$ can be used,
resulting in the DM contrast,
\begin{eqnarray}\label{crbc1}
C(\vec{\rho})=\frac{\Delta n}{n_{ob}}\Bigg[\bigg(z_f -
h_1(\vec{\rho})\bigg) {{\bigtriangledown^{2}}h_{1}(\vec{\rho})} +\nonumber\\
\bigg(z_f+|h_2(\vec{\rho})|\bigg){{\bigtriangledown^{2}}|h_{2}(\vec{\rho})|}\Bigg],
\end{eqnarray}
with $\frac{1}{\sqrt{A}}{\sum_{\vec{q}}}
h(\vec{q})sin(\vec{q}\cdot\vec{\rho})q^2 = - \bigtriangledown
^{2}h(\vec{\rho})$. From this expression we note that DM contrast is
proportional to the object's local curvature
($\nabla^{2}{h}\equiv\kappa$). Therefore, the flat glass-substrate
is not visualized with DM. The limit of validity of Eq.
(\ref{crbc1}) can be checked by measuring the linear region of
$C\times z_f$. For RBC this limit is $z=\pm 2\mu$m in relation to
the middle plane (minimum contrast plane). Since the roughness of
the interface $u(\vec{\rho},t)$ can have large spatial wavenumbers
q, Eq. (\ref{ap2g}) must be used to analyze height fluctuations of
the RBC surface-membranes\cite{giuseppe,livia}.

It is important to stress that the introduction of the
surface-membranes height coordinates, $h_{1}(\vec{\rho})$ and
$h_2(\vec{\rho})$, into the phase factors $((z_{f}-h_{1/2})q^2/2k)$
in Eq. (\ref{ap3g}), that allowed us to achieve total $3$D imaging.
None of the previous approaches using defocusing techniques had
considered these important factors before. The coupling between
$h(\vec{\rho})$ and $\nabla^{^2}h(\vec{\rho})$, as shown in Eq.
(\ref{crbc1}), is responsible for the high axial resolution achieved
by Defocusing Microscopy, as will be shown later.

To obtain RBC thickness profile and volume, one needs to measure the
time-average cell contrast at two different defocus distance:
$<C_1>$ (at $z_{f_1}$) and $<C_2>$ (at $z_{f_2}$). Subtracting,
pixel by pixel, the contrast image $<C_2>$ from $<C_1>$,
\begin{eqnarray}
<C_2> - <C_1>=\frac{\Delta n}{n_{ob}}(z_{f_2} -
z_{f_1})\bigtriangledown^{2}H,
\label{contrasteduasE2}
\end{eqnarray}
where $H(\vec{\rho})=h_1(\vec{\rho}) + |h_2(\vec{\rho})|$ is the
cell thickness. Performing a Fourier transform on Eq.
\ref{contrasteduasE2}, the laplacian term gives rise to a
$(-q^2)(h_{1}(\vec{q})+|h_2|(\vec{q}))$ term, which divided by
$-q^2$ and performing an inverse Fourier Transform,
\begin{eqnarray}\label{perfilaltura2}
H=\frac{n_{ob}}{\Delta
n(z_{f_1}-z_{f_2})}{\mathcal{F}}^{-1}\Bigg(\frac{{\mathcal{F}}\{<C_2>-<C_1>\}}{-q^{2}}
\Bigg).\nonumber\\
\end{eqnarray}
If the values for $z_f$, $n_{ob}$ and $\Delta n$ are known, the
thickness profile map of the observed cell is obtained. In addition,
if the pixel area value is known, the cell volume $V= A_{pixel}
\times H(\vec{\rho})$ is calculated. The mean thickness profiles for
the $25$ analyzed cells in hypotonic, isotonic, and hypertonic media
are shown in the supplemental material \cite{sm}, and the mean
volume values are shown in Table $1$ and are in accordance with
those reported by others techniques
\cite{rappaz2008,popescu2011,evans,laser2009}.

Using the thickness profile $H(\vec{\rho})$ and defining
$\delta(\vec{\rho})= h_1 - |h_2|$ as the asymmetry between the RBC
surface-membranes $h_1$ and $h_2$, we then have,
\begin{eqnarray}\label{crbc4}
h_1(\vec{\rho})=\frac{H(\vec{\rho})+\delta(\vec{\rho})}{2}\;\;\;\;\;\;\;\;h_2(\vec{\rho})=\frac{-H(\vec{\rho})+\delta(\vec{\rho})}{2}.\nonumber\\
\end{eqnarray}
For $z_{f}=0$, Eq. (\ref{crbc1}) can be rewritten as,
\begin{eqnarray}\label{crbc3}
\nabla^2\delta+\frac{\nabla^2H}{H}\delta=-\frac{2n_{ob}}{\Delta n
H}C(0,\vec{\rho}),
\end{eqnarray}
which is a non-homogeneous Helmholtz equation with variable
coefficients, that can be numerically solved for
$\delta(\vec{\rho})$ with the initial condition
$\delta(\vec{\rho})=0$. Since
$\frac{\nabla^2H(\vec{\rho})}{H(\vec{\rho})}$ is a well-behaved
function, the convergence of the fitting procedure is very robust.
From the returned $\delta(\vec{\rho})$ the total $3$D imaging is
obtained, as shown in Fig.\ref{solida} (a-f) (see also movies
$S1-S3$ in supplementary materials \cite{sm}). The two extreme
experimental points in the border of each interface are smoothly
connected by a circumference, because diffraction at the border
distorts the actual profile.

In Fig.\ref{solida} (g-i), the average angular profiles for each group of RBCs subject to different osmotic pressures are
displayed and the differences between the upper and lower membranes
are evidenced. For
hypotonic solution (Fig.\ref{solida}g), the upper
membrane is inflated due to water entrance and the lower membrane
remains attached to the glass-substrate. In the isotonic case
(Fig.\ref{solida}h), it is seen that due to cell adhesion the lower
membrane has a flatter height profile than the upper membrane, and for the
hypertonic case (Fig.\ref{solida}i), the
separation distance between the lower and the upper membrane
decreases, and the RBC exhibits a more pronounced dimple in the
center region for both membranes. For all cases the adhesion contact region between the lower membrane and the
glass-substrate is located (Fig.\ref{solida}b and c): isotonic $\rho=2.5\mu$m and hypertonic $\rho=3\mu$m. Since the height
profile and the pixel area are known, the cell
surface area can be calculate as
$A=\displaystyle{\int\sqrt{1+\bigg(\frac{\partial h_{1/2}}{\partial
x}\bigg)^2+\bigg(\frac{\partial h_{1/2}}{\partial y}\bigg)^2}dxdy}$. From the
surface area and volume, we calculate the sphericity index
$(\chi)$\cite{canham1968}. All values are shown in Table $1$ and are in accordance with
those reported by other techniques \cite{evans,canham}.

\begin{figure*}
\centering\includegraphics[width=16cm]{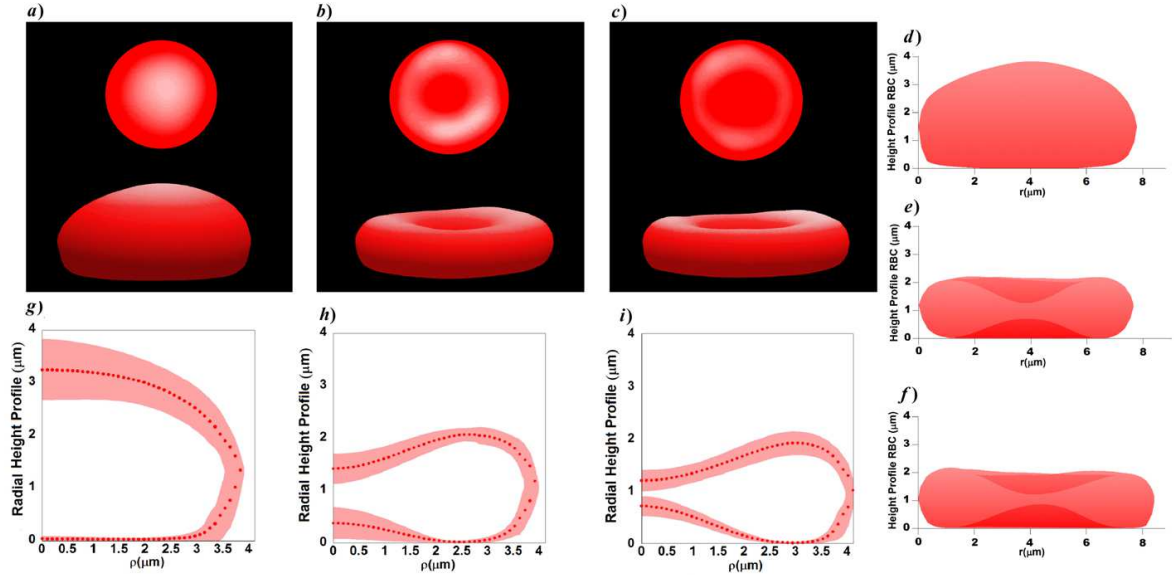}
\caption{\label{solida} Total $3$D Imaging of red blood cells in
(a/d) hypotonic (200 mOsm/Kg), (b/e) isotonic (300 mOsm/Kg) and
(c/f) hypertonic (400 mOsm/Kg) solutions (Multimedia view). (g-i)
Radial height profiles over all analyzed cells. The shaded areas
represent the standard deviations of the average over the $25$
RBCs.}
\end{figure*}

The DM $3$D imaging method is independent of the chosen reference
system, and thus, it can be efficiently used to image red blood
cells which are flowing through a glass capillary or a micro-fluidic
device. To avoid moving the objective, one can use a beam-splitter
to split the images and use two cameras focused in different
positions to collect them. The relation between objective defocusing
$(Z_{o})$ and camera defocusing $(Z_c)$ is $Z_c = M^2 Z_{o}$, where
M is the magnification of the objective. For a $M=100 X$ objective,
the defocusing effect of a displacement of $1\mu$m is obtained by
displacing the camera by 1$cm$. Therefore, by using two cameras with
an optical path difference of $1$ cm, the images recorded by each
camera will correspond to the contrasts $C_1$ and $C_2$ of Eq.
(\ref{contrasteduasE2}), where $z_{f1}-z_{f2}=1\mu$m. The DM
technique can also be used for total $3D$ imaging of high motility
cells and to obtain information about the upper and lower membranes,
the last one in dynamical contact with the substrate.

\begin{table*}
\caption{\label{tab:table4}Data on RBC osmolality, volume, surface
area, sphericity index and RBC refractive index obtained from the
full 3D imaging.}
\begin{ruledtabular}
\begin{tabular}{cccccc}
Osmolality&Radius$(\overline{R}_M)$&\mbox{Volume $(\overline{V}_M)$}&\mbox{Surface Area$(\overline{A}_M)$}&\mbox{$\overline{\chi}$}&\mbox{$n_{rbc}$}\\
(mOsm/Kg)&\mbox{($\mu$m)}&\mbox{($\mu$m$^3$)}&\mbox{($\mu$m$^2$)}&\mbox{$--$}&\mbox{$\lambda=610$nm}\\
\hline
200$\pm2$&3.85$\pm0.23$&129$\pm15$&134$\pm11$&0.92$\pm0.02$&1.378$\pm0.003$\\
300$\pm2$&3.92$\pm0.17$&94$\pm8$&130$\pm9$&0.77$\pm0.01$&1.391$\pm0.003$\\
400$\pm2$&4.12$\pm0.20$&\mbox{91$\pm8$}&\mbox{139$\pm11$}&0.70$\pm0.01$&\mbox{1.400$\pm0.005$}\\
\end{tabular}
\end{ruledtabular}
\end{table*}

\begin{figure}
\centering\includegraphics[width=6cm]{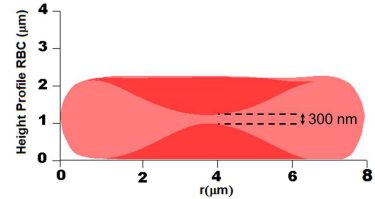}
\caption{\label{fig4} Total $3$D image of a RBC in a hypertonic
solution (400 mOsm/Kg). The surface-membranes are clearly imaged,
indicating that the axial resolution in this case is better than
$300$nm (Multimedia view).}
\end{figure}

Defocusing Microscopy axial resolution depends on the sensitivity of
image contrast measurement, optical contrast of the phase object,
and mean curvature of the surfaces considered. For example, using
our Eq. (\ref{crbc1}) at $\rho=0$ and $z_f= h_1$, we obtain,
\begin{eqnarray}\label{axial_1}
C(0,h_1)= \frac{\Delta n}{n_{ob}}(h_1 + |h_2|)\nabla^2 h_{1},
\end{eqnarray}
and for $z_f=-|h_2|$,
\begin{eqnarray}\label{axial_2}
C(0,h_2)= \frac{\Delta n}{n_{ob}}(-h_1 - |h_2|)\nabla^2 h_{2},
\end{eqnarray}
such that $\Delta C=C(0,h_1)-C(0,h_2)$ then,
\begin{eqnarray}\label{axial_6}
\Delta C= \frac{\Delta n}{n_{ob}}(h_1 + |h_2|)(\nabla^2
h_{1}+\nabla^2 h_{2}).
\end{eqnarray}
For $\rho=0$, $d=h_1 + |h_2|$ is the axial distance between the two
surface-membranes in the RBC center, as shown in Fig.
\ref{hemacia}a, and since for RBCs
\begin{eqnarray}\label{axial_3}
\nabla^2 h_{1}\simeq\nabla^2 h_{2}=\kappa,
\end{eqnarray}
where $\kappa$ is the curvature of the surfaces then,
\begin{eqnarray}\label{axial_4}
\Delta C= \frac{2\Delta n}{n_{ob}}d\;{\kappa}.
\end{eqnarray}
The minimum axial distance $d_{min}$ is thus given by,
\begin{eqnarray}\label{axial_5}
d_{min}= \frac{n_{ob} \Delta C_{min}}{2 \Delta n {\kappa}}.
\end{eqnarray}
Using $\Delta C_{min}=10^{-2}$ as the contrast sensitivity, an oil
immersion objective $(n_{ob}=1.5)$ and the typical values for
isotonic RBCs, $\Delta n$=0.06, and $\overline{\kappa} =
0.7\;\mu$m$^{-1}=0.7\times10^{-3}$ nm$^{-1}$, the minimum axial
distance is approximately $180$ nm, sufficient to resolve the axial
separation of the RBCs surface-membranes, for all solution
osmolalities studied. In Fig.\ref{fig4} a total $3$D image of a RBC
in a hypertonic solution is presented, showing that we are able to
resolve the distance between the two membrane surfaces at the cell
center, which is $300$ nm and therefore confirms the estimates above
(see movie $S4$ in supplementary materials \cite{sm}).

In order to obtain $H(\vec{\rho})$ we need the value for $\Delta n$,
which is basically defined by the RBC hemoglobin concentration. A
good assessment of the red cell refractive index ($n_{rbc}$) can be
obtained by measuring the distance between the minimum contrast
plane to the glass coverslip plane, which is $|h_{2_{m}}|$ in the
RBC rim position. The layout of the refractive index experiment is
seen in Fig.\ref{hemacia}a, where polystyrene beads of diameter
$d=0.198\;\mu$m are deposited on the glass coverslip, such that the
location of the glass slide is determined from a z-scan throughout
the bead defocused images. The minimum bead contrast occurs when the
objective focal plane coincides with its diameter, from where we
move the objective until the RBC minimum contrast plane (largest
diameter plane). To be more quantitative, one can select images
around the bead and around the RBC as the objective focal plane is
scanned. The recorded images must contain the bead and the RBC
completely, as shown in Fig.\ref{hemacia}(b-d). Since there is no
light absorption, the spatial average intensity of the complete
image is equal the background intensity,
$<I>_{\vec{\rho}}=I_0=(1/N_p)\sum_{i}\ I_i$,\;\;\; where $N_p$ is
the total number of pixels of the image and $I_{i}$ is the intensity
of each pixel. In order to define the minimum contrast position we
calculate the standard deviation of the spatially averaged intensity
$I$ for each frame and divide it by $I_0$, which corresponds to
$\sqrt{<I^2>-<I>^2}/{<I>}=\sqrt{<\Delta C^2>_{\vec{\rho}}}$, with
$<I^2>=(1/N_p)\sum_{i}\ (I_i)^2$. In Fig.\ref{hemacia}e, the
behaviour of $\sqrt{<\Delta C^2>_{\vec{\rho}}}$ as a function of
$z_f$ is shown, from which the minimum values of $\sqrt{<\Delta
C^2>_{\vec{\rho}}}$ for a RBC ($z_f=0$) and of the polystyrene bead
($z_f=z_{bead}$) are clearly determined. Hence,
$|h_{2m}|=|z_{bead}|+|r_{bead}|=|z_{bead}|+0.099\mu m$. The obtained
values for the RBC refractive ($\lambda=0.610\mu$m) are shown in
Table $1$, in agreement with results obtained using different
techniques \cite{rappaz2008,laser2009}.  As expected, due to water
entrance into the cell, for the hypotonic solution $n_{rbc}$ is
lower than in the other cases.

In conclusion, in this paper we present a novel methodology to
obtain total $3$D imaging of phase objects using a defocused
bright-field optical microscope. We apply the method to RBCs subject
to different solution osmolalities and determine the cell shapes and
surface-membranes deformation due to adhesion to the
glass-substrate. Strikingly, our results show that the Defocusing
Microscopy axial resolution for RBCs can be smaller than 300nm,
allowing a clear imaging of both surface-membranes. In addition, we
obtain the cell average refractive index, surface area, volume and
sphericity index over a group of $25$ red blood cells under
different osmotic conditions, showing that DM technique can be used
for monitoring certain pathologies. Since DM does not require the
use of any extra optical element, the technique could be easily
adopted by non-specialists.

\vspace{0.2cm}
\textbf{Acknowledgements}

We would like to acknowledge the financial support from the
Brazilian Agencies CNPq, FAPEMIG, PRONEX-FACEPE, Instituto Nacional
de Fluidos Complexos e Aplica\c{c}\~oes (INFCx) and FAPEAM. The
authors also thank Michael O'Carroll for useful discussions.

\nocite{*}
\bibliography{aipsamp}

\begin{thebibliography}{99}

\bibitem {zernicke} F. Zernike,  Physica {\bf 9}, 686 (1942).

\bibitem {nomarski} G. Nomarski,  J. Phys. Radium {\bf 16}, 9 (1955).

\bibitem {nugent1} D. Paganin and K. A. Nugent, Phys. Rev. Lett. {\bf 80}, 2586 (1998).

\bibitem {nugent2} E. D. Barone-Nugent, A. Barty, K. A. Nugent,  J. Micros. {\bf 206}, 194 (2002).

\bibitem{rappaz2008} B. Rappaz et al., Cytometry {\bf 73A}, 895 (2008).

\bibitem {popescu2008} G. Popescu et al., Blood Cells Mol. Dis.  {\bf 41}, 10 (2008).

\bibitem {kononenko} V. L. Kononenko,  Wiley-VCH Verlag GmbH Co. KGaA. 155 (2011).

\bibitem{giuseppe} G. Glionna et al., Appl. Phys. Lett. {\bf 15 94}, 193701 (2009).

\bibitem{livia} L. Siman et al., submitted to publication (2014).

\bibitem{popescunature} T. Kim et al., Nature Photonics, Advance Online Publication, 19 January (2014).

\bibitem {teague} M. R. Teague,  J. Opt. Soc. Am. {\bf 73}, 1434 (1983).

\bibitem {wolf} M. Born and E. Wolf, ``Principles of Optics'', Cambridge University Press, New York (1999).

\bibitem{goodman} J. W. Goodman,``Introduction to Fourier Optics,'' McGraw-Hill Co., Inc.(2002).

\bibitem{sm} See supplementary material at () for thickness profile results, movies of red cells $3$D imaging, as well as information on the experimental set up, materials and procedures.

\bibitem {leo} L. G. Mesquita, U.  Agero, and O. N. Mesquita, Appl. Phys. Lett. {\bf 88}, 133901 (2006).

\bibitem{bira2003} U.  Agero et al., Phys. Rev. E {\bf 67}, 051904 (2003).

\bibitem{coelho2005} J. C. Neto et al.,Exp. Cell Res. {\bf 303 (2)}, 207 (2005).

\bibitem {coelho2007} J. C. Neto et al., Biophys. J. {\bf 91}, 1108 (2006).

\bibitem{popescu2011} Y. Park et al., Phys. Rev. E {\bf 83}, 1051925 (2011).

\bibitem{rappaz2009} B. Rappaz et al. Blood Cell, Mol. Dis. {\bf 42}, 1 (2009).

\bibitem{evans}E. Evans and Y. C. Fung,  Microvascular Research {\bf 4}, 335 (1972).

\bibitem{laser2009} A. L. Yusipovich et al., J. Appl. Phys. {\bf 105}, 102037 (2009).

\bibitem{canham1968}{P. B. Canham and A. C. Burton}, Circulation Research {\bf 22}, 405 (1968).

\bibitem{canham} P. B. Canham, J. Theor. Biol. {\bf 26}, 61 (1970).

\end{thebibliography}

\end{document}


\preprint{AIP/123-QED}

\title[]{Supporting Information: \\Total 3D imaging of phase objects using defocusing microscopy: application to red blood cells}

\author{P. M. S. Roma$^{1}$, L. Siman$^{1}$, F. T. Amaral$^{2}$, U. Agero$^{1}$ and O. N. Mesquita}
\email[E-mail me at: ]{mesquita@fisica.ufmg.br}

\address{$^{1}$Departamento de F\'\i sica, $^{2}$Programa de P\'os Gradua\c{c}\~ao em
Engenharia El\'etrica, Universidade Federal de Minas Gerais, Caixa
Postal 702, CEP 31270-901, Belo Horizonte, MG, Brazil.}
\email[E-mail me at: ]{mesquita@fisica.ufmg.br}

\maketitle

\subsection{Defocusing Microscopy Instrumentation}
Experiments were conducted on inverted microscopes operating in
bright field (Nikon TE$300$ and Nikon Eclipse TI-E) with oil
immersion objectives (Nikon Plan APO DIC H, $100$X, $1.4$ NA and
Nikon Plan APO DIC H, $100$X, $1.49$ NA; Nikon, Melville, NY).
Images of RBCs were captured with a CMOS camera (CMOS 642M) at a
typical frame rate of $300$fps. The defocusing distance was
controlled by a piezoelectric nanoposition stage (P$563-3$CD, Physik
Instrumente (PI) GmbH Co, Karlsruhe, Germany) coupled to the
microscope stage. Perfect Focus System (Nikon) was used to stabilize
the focus. A red filter ($\lambda = 0.61\mu$m) was added into the
microscopes optical paths in order to avoid absorbing effects by
RBCs.

\subsection{Cell Sample Preparation}

Briefly, approximately $0.5\mu$L of fresh blood is collected from
healthy laboratory personnel, and suspended in a $1.5$ml solution of
phosphate-buffered saline (PBS) (pH 7.4, NaCl 150 mM) with
$1.0$mg/mL of bovine serum albumin (Sigma Aldrich, St. Louis, MO)
\cite{leo}. This solution has a refractive index of
$n_{solution}=1.333\pm0.001$ and represents our isotonic solution
(NaCl 0.9\% - 300 mOsm/kg). We adjusted the NaCl concentrations to
0.6\% and 1.2\% to obtain the desired osmolality corresponding to
hypotonic (200 mOsm/kg) and hypertonic (400 mOsm/kg) solutions,
respectively. The osmolality of each solution was measured with a
freezing-point depression osmometer (Advanced Model 3250
Single-Sample Osmometer). For each experiment, a $250\mu$L volume of
the RBCs suspension was transferred to a microscope cover glass
chamber and covered with a coverslip to avoid water evaporation. For
refractive index measurements, the same volume of the RBC suspension
was transferred to a pre-treated cover glass with polystyrene beads
(Polysciences, Inc, Warrington, PA, USA) of $0.198 \mu m$. Samples
were made immediately before experiments and left still on the
microscope stage for approximately $15$ minutes to allow cell
precipitation and adhesion. On average, each experiment lasts for
two hours. All experiments were conducted at room temperature
$(\sim27\,^{\circ}\mathrm{C})$.

\subsection{Thickness Profile}
\begin{figure*}
\centering\includegraphics[width=15cm]{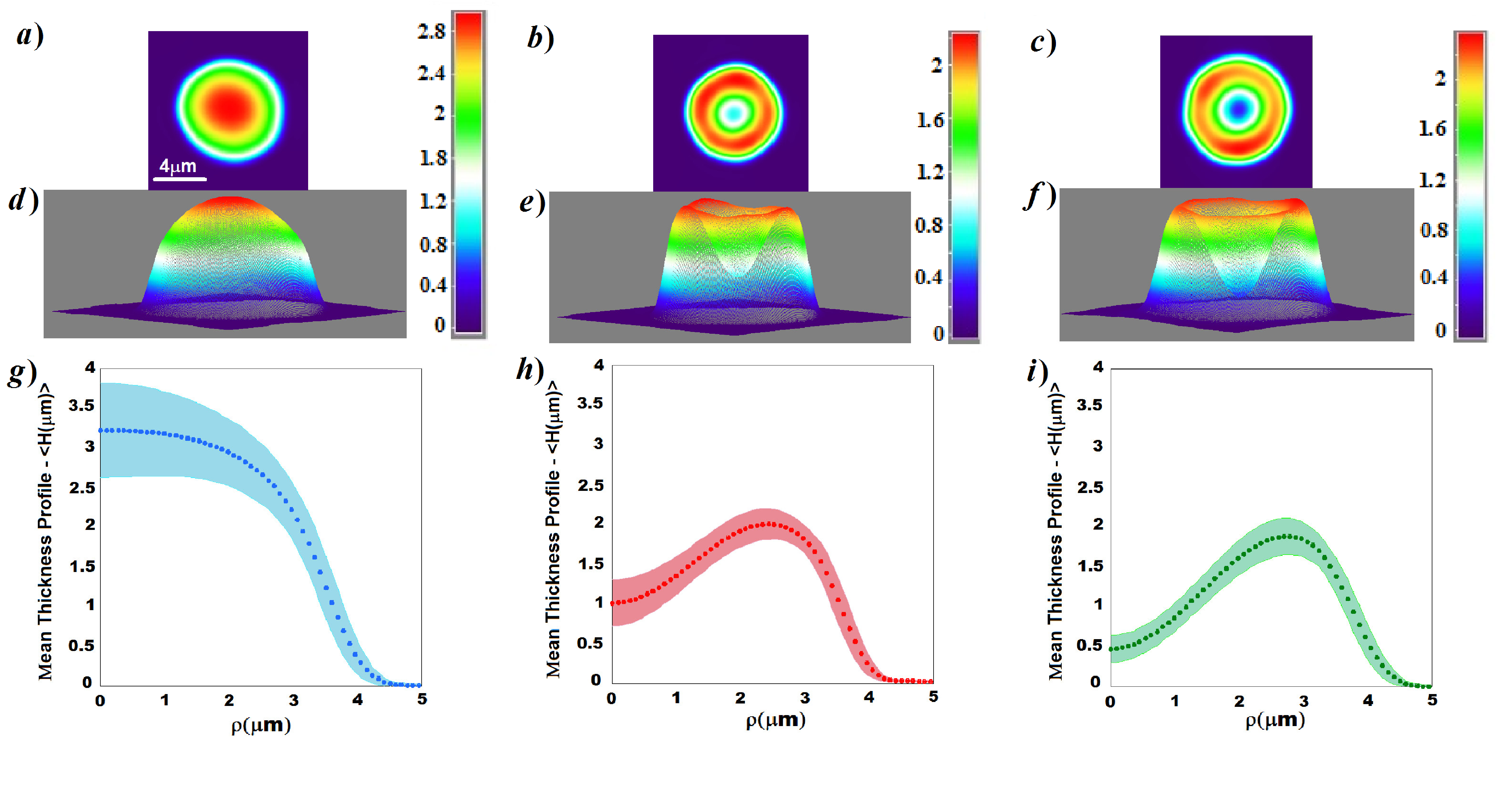}
\caption{\label{espessura} Thickness profile map of red blood cells
immersed in a (a) hypotonic (200\;mOsm/kg), (b) isotonic
(300\;mOsm/kg), and (c) hypertonic (400\;mOsm/kg) solutions, and its
respective $3$D thickness reconstruction (d - f).(g - h) Mean
thickness radial profile for the $25$ analyzed cells in each
solution. The shaded areas represent the standard deviations.}
\end{figure*}

The mean thickness results for $25$ cells in hypotonic, isotonic,
and hypertonic solutions are shown in Fig. \ref{espessura}. In Fig.
\ref{espessura}(a - c), a red cell thickness map immersed in each
situation is presented and in (d - f) the respective $3$D thickness
images are shown. The 3D thickness image is obtained using ImageJ
Interactive 3D Surface Plot plugin software package (Rasband WS.
ImageJ, U.S. National Institutes of Health, Bethesda, Maryland, USA,
imagej.nih.gov/ij/, 19972012)\cite{image} over the thickness map.
The mean thickness profile and cell radius obtained is in agreement
with previous findings\cite{popescu2008,popescu2011,rappaz2009}. In
Fig.\ref{espessura}(g - i) the average radial profile over each set
of $25$ RBCs are presented, indicating that the medium osmolality
influences the cell shape. In the hypotonic medium
(Fig.\ref{espessura}g, the RBCs are swollen due to water influx,
losing its discoid shape exhibited in isotonic solution (Fig.
\ref{espessura}h). In the hypertonic case (Fig.\ref{espessura}i),
the RBCs suffers water efflux, such that the dimpled region in the
center becomes more pronounced.

\subsection{Image processing and numerical procedures}

Images captured with a CMOS camera are later processed, because even
under uniform illumination the background gray level of each pixel
is not uniform. We used a correction procedure as established on the
ImageJ Information and Documentation Portal website. This procedure
consists of filming the area of interest without RBCs, with uniform
illumination, and measure the gray level of each pixel. Later the
gray level of each pixel of the images with the RBCs are divided by
the background gray level, measured previously. Therefore, any pixel
inhomogeneity is eliminated. All captured images were analyzed using
NIH-ImageJ software package, Java plugins and algorithms in Matlab
(The Mathworks Inc., Natick, MA) developed in our group to perform
the complete 3D reconstruction. The model fitting was implemented
using Matlab's lsqcurvefit function, which is based on least squares
method. The asymmetry parameter was adjusted in order to fit the
analytical expression presented on equation (8) to the experimental
data. The least squares method requires a starting value to the
fitting parameter. We set 0 as the starting value to $\delta$ for
the tests after being observed that the convergence of the algorithm
was satisfactory for this initial guess. However, we tested
different initial guesses of the parameter $\delta$, which also
yielded optimal solutions and returned the same result for $\delta
(\vec{\rho})$, showing that the fitting method is robust and
appropriate to be applied to the problem of $3$D imaging of phase
objects. Then for all subsequent fittings we started from $\delta
(\vec{\rho})=0$. The stop criteria used on the optmization routine
was the sum at squared error less or equal than $0.1\%$ of the
experimental data variance. In average, after $10$ iterations the
algorithm converged.

\subsection{Video of RBC Total $3$D Imaging}

Movie S1. The movie file \emph{Hypotonic.mov} illustrates the total
$3$D Imaging of a RBC immersed in a hypotonic solution (200
mOsm/Kg). The upper surface-membrane is inflated due to water influx
and presents a nearly spherical profile, while the lower
surface-membrane becomes flat and almost completely attached to the
glass-substrate.

Movie S2. The movie file \emph{Isotonic.mov} illustrates the total
$3$D Imaging of a RBC immersed in an isotonic solution (300
mOsm/Kg). The cell has a biconcave shape and the lower
surface-membrane has a flatter profile than the upper one, due to
adhesion to the glass-substrate.

Movie S3. The movie file \emph{Hypertonic1.mov} illustrates the
total $3$D Imaging of a RBC immersed in a hypertonic solution (400
mOsm/Kg). Due to water efflux, the center separation distance
between the lower and the upper surface-membranes is decreased and a
more pronounced dimple in the center is seen.

Movie S4. The movie file \emph{Hypertonic2.mov} illustrates the
total $3$D Imaging of a RBC immersed in a hypertonic solution (400
mOsm/Kg), with a distance between the two surface-membranes at the
cell center of 300nm. As can be seen, we are able to obtain a more
complete assessment of the shapes and deformations of the two
surface-membranes for different angles of view.

All movies are rotations of a static RBC which was first rotated in
z and then in x axis for 360$^{\circ}$.

\nocite{*}
\bibliography{aipsamp}